\documentstyle[twoside,fleqn,espcrc2]{article}

\def\CM{{\cal{M}}}
\def\str{{\rm str}}
\def\Mhat{{\hat M}}

\title{Applications of partially quenched chiral perturbation theory
       }

\author{Maarten Golterman\address{Department of Physics, Washington
University,
        St. Louis, MO 63130, USA.}
        and
        Ka Chun Leung\address{Institute of Theoretical Physics III, 
        University of Siegen, 
        57068 Siegen, Germany.}
        \thanks{presenter at conference}
        }

\begin{document}

\begin{abstract}
Partially quenched theories are theories in which the
valence- and sea-quark masses are different.
Here, we will discuss the nonanalytic
one-loop corrections to some physical quantities,
using partially quenched chiral perturbation theory. 
In particular, we will focus on the results for Goldstone boson masses 
to illustrate the general features of our calculation. 

\end{abstract}

\maketitle
\section{Partially Quenched Chiral Perturbation Theory (PQChPT)}
We will introduce PQChPT
following the formulation in ref.\ \cite{BG}. 
For simplicity, we will restrict ourselves 
to the case in which  there are 3 
quarks $u$, $d$ and $s$, with masses $m_u$, $m_d$ 
and $m_s$ respectively, and a set of $N$ 
degenerate quarks $q_S$ (with common mass $m_S$) 
(the arguments can easily be generalized to 
other cases). We wish to consider 
lattice computations in which only the 
$N$ quarks $q_S$ contribute as sea quarks. 
Hence, we add three additional {\it bosonic} 
quarks $\tilde u$, 
$\tilde d$ and $\tilde s$ with the same other 
quantum numbers and masses as their counterparts $u$, $d$ and 
$s$, which will cancel the sea-quark effects caused
by the $u$, $d$ and $s$ quarks.

If one assumes that, as in unquenched
QCD, the strong interactions between 
these quarks are responsible for chiral 
symmetry breaking and confinement, one expects
Goldstone mesons  corresponding to the breakdown of 
the graded chiral symmetry group
$SU_L(3+N|3)\times SU_R(3+N|3)$ to
$SU_{L+R}(3+N|3)$  by the vacuum state 
(quark masses can be incorporated like in full ChPT).  
However, the spectrum of Goldstone mesons is 
enlarged relative to full QCD, and is described 
by the $(6+N)\times (6+N)$ hermitian matrix 
\begin{equation}
\Phi=\pmatrix{\phi&\chi^\dagger\cr\chi&{\tilde{\phi}}\cr}\ ,
\end{equation}
where $\phi$ is the $(3+N)\times (3+N)$ matrix of ordinary 
mesons made from the $3+N$
ordinary quarks and antiquarks, $\tilde \phi$ 
is the corresponding
$3\times3$ matrix for bosonic-quark mesons, 
and $\chi$ is a $3\times
(3+N)$ matrix of fermionic
mesons made from a bosonic quark and an ordinary antiquark.
One can then construct, using the unitary field $\Sigma$ 
defined through 
\begin{equation}
\Sigma \equiv \exp{(2i\Phi/f_p)}\ ,
\end{equation}
and the diagonal  mass matrix $\hat M$ with entries
\begin{equation}
m_u ,m_d, m_s, m_S,\cdots,m_S,m_u,m_d,m_s
\end{equation}
on the diagonal (where $m_S$ occurs $N$ times), 
the $O(p^2)$ Euclidean partially quenched chiral Lagrangian as:
\begin{eqnarray}
{\cal L}&=&{f_p^2\over 8}\; \str\left( \partial_\mu \Sigma \partial_\mu
\Sigma^\dagger \right)\nonumber \\
&&-{f_p^2\mu_p\over 4}\;\str\left(\Mhat\Sigma+\Mhat\Sigma^\dagger\right)
\nonumber \\
&&+{m^2_0\over
6}\; \Phi^2_0 +{\alpha\over 6}
\left( \partial_\mu \Phi_0 \right)\left(\partial_\mu \Phi_0 \right), 
\label{L}
\end{eqnarray}
where {\rm str} denotes the supertrace, and 
$\Phi_0 \equiv{\rm str}(\Phi)$ 
is invariant under the chiral symmetry group, and 
is called super-$\eta'$. $f_p$ is the tree-level 
weak decay constant; 
$m_0$ and $\alpha$ are parameters introduced through 
the $\Phi_0$ terms. 

\section{Interplay between $m_0$ and $m_S$}
The $O(p^2)$ two-point functions can be read off  
from the Lagrangian eq.~(\ref{L}). In particular, let us consider 
the $\eta'$ two-point function 
(in the mass-degenerate case $m=m_u=m_d=m_s$, 
and with $\alpha=0$ for simplicity). 
The two-point function reads  \cite{BG}
\begin{eqnarray}
\langle \eta' \eta' \rangle =\!\!\!\!\!\!\!\!\!
&&{1\over p^2 +m^2_\pi}\nonumber \\
&&+{-m^2_0 \over (p^2 +m^2_\pi)^2} 
{p^2 +M^2_{SS} \over p^2 +M^2_{SS}+Nm^2_0/3}\ , \label{tp}
\end{eqnarray}
where the tree-level pion mass-squared $m^2_\pi =2\mu_p m$, and 
$M^2_{SS}=2\mu_p m_S$. 

A number of comments is in order. 
First, the combination $m^2_{\eta'}=M^2_{SS}+Nm^2_0/3$ is the
mass-squared of 
the super-$\eta'$  
at tree level. 
Second, in the situation in which the sea quarks are integrated 
out (hence their dynamical effects are absent), which 
mathematically corresponds to setting $N=0$ or taking
$M_{SS}$ much larger than other relevant energy scales 
({\it i.e.} external momenta, $m_0$ and  $m_\pi$), 
the last factor in eq.~(\ref{tp}) turns into unity and the two-point 
function degenerates into the form  predicted by fully quenched 
ChPT, with the double-pole term which gives rise to ``quenched 
chiral logarithms." In the partially quenched case, the double pole
is also present for $M_{SS}\ne m_\pi$. 
The sea-quark dynamics is encoded in this last factor in which 
the interplay of the two scales $M_{SS}$ and $m_0$ is evident. 
We will show below in an explicit example 
how this interplay occurs in physical quantities.  
Third, one can also envision integrating 
out the super-$\eta'$ 
when  $m_0$ is much larger than the other mass scales. This is 
the limit considered in ref.\ \cite{Sharpe}. 
Finally, in the unquenched theory, which is obtained by setting 
$N=3$ and $m_S=m$, we obtain for the two-point function  
\begin{equation}
\langle \eta' \eta' \rangle_{full} ={1\over p^2+m^2_\pi +m^2_0}\ ,
\end{equation}
in which the double pole has disappeared.   
Again, the $\eta'$ decouples from the theory if, in addition,  
$m_0$ is heavy compared to other energy scales. 
This corresponds to conventional ChPT 
in which $\eta'$ is not explicitly represented. 

We thus observe that PQChPT contains different mass 
scales (all assumed to be below the chiral symmetry breaking 
scale $\Lambda_p$): that of the valence quarks and those introduced 
by  partial quenching, {\it i.e.} $m_0$ and $m_S$.
In particular, the ratio of $M_{SS}$ and $m_0$ is 
arbitrary. One can develop 
PQChPT systematically and it ``interpolates" between 
the fully quenched and unquenched theories. 
It allows us to assess the situation 
in which $M_{SS}$ is of the order $m_0$, which is 
likely to be the case in typical
partially quenched lattice computations.  
We have calculated the nonanalytic one-loop corrections for
the chiral condensate, weak decay
constants, Goldstone-boson masses,
$B_K$, all  with nondegenerate valence-quark masses,  
and the $K^+ \rightarrow \pi^+ \pi^0$ decay
amplitude with degenerate valence-quark masses. 
In the following, we will focus only on the 
one-loop mass predictions in the degenerate valence-quark 
limit. We refer to ref.\ \cite{GL}
for further discussion of other quantities. 

\section{Method}
The complete $O(p^4)$ predictions of PQChPT contain contributions 
from  
$O(p^4)$ operators whose associated coefficients are 
largely unknown. To further complicate the matter, 
there are also one-loop contributions, whose 
coefficients contain unknown $\eta'$ coupling constants,  
proportional to $\log{m^2_{\eta'}}$ coming from 
$\Phi_0$ tadpoles, which thus 
give rise to a complicated dependence on $m_S$ and $m_0$. 
To simplify and still make useful predictions, we 
adopt a procedure in which only the  
loop corrections nonanalytic in the valence-quark 
mass are kept, and 
each fixed value of the sea-quark mass $m_S$ corresponds 
to a different partially quenched theory. This then allows 
us to ignore contributions from $\Phi_0$ tadpoles.
We also expand the results in $m^2_\pi /m^2_{\eta'}$ 
for further simplification. This preserves the 
property that PQChPT predictions give the fully quenched results
for $M_{SS}\to\infty$, or the unquenched result for $M_{SS}=m_\pi$
(and $m_0\to\infty$),
when comparing the nonanalytic one-loop corrections. 
As the fully quenched and unquenched results already existed, 
this offers a nontrivial check of our results. 

The comparison between our results \cite{GL} 
and those of ref.\ \cite{Sharpe}, however,  
is actually more delicate, and deserves some more explanation. 
In our case, we keep the super-$\eta'$, whereas in ref.\ \cite{Sharpe} the 
super-$\eta'$ is integrated out. One would therefore in general expect 
that in order to ``match" the two theories, we would need 
to adjust the bare parameters. For the quantities considered 
in ref.\ \cite{GL}, it turns out that, at one loop, all nontrivial adjustments 
come from $\Phi_0$-tadpole contributions. Since we did not 
include such contributions, we conclude that, 
for the comparison, no adjustment is needed.  

Eventually, one can obtain 
numerical estimates of one-loop corrections, 
taking, for example, a value like $700$ MeV 
or $1$ GeV (as in refs.\ \cite{GL,GLquenched}) 
for the cutoff $\Lambda_p$ 
in the chiral logarithms.  The difference should gives an idea of  
the size of the unknown contributions from $O(p^4)$ operators 
and $\Phi_0$ tadpoles.

\section{Physical Predictions}
The mass-squared for the pions with degenerate valence quarks 
at one loop then reads
\begin{equation}
\left[ m^2_\pi \right]_{\rm 1-loop} =m^2_\pi \left( 
1-{2\over 3(4\pi f_p)^2}I \right)\ ,
\end{equation}
where 
\begin{equation}
I=B+\left(B+C\right)\log{m^2_\pi \over \Lambda^2_p}\ , \label{full}
\end{equation}
\begin{equation}
B={(m^2_\pi -M^2_{SS})(m^2_0 -\alpha m^2_\pi)\over 
  m^2_\pi -M^2_{SS} +{N\over 3}(\alpha m^2_\pi -m^2_0)}\ ,
\end{equation}
and 
\begin{equation}
C\!\!=\!\!-m^2_\pi {\alpha (m^2_\pi -M^2_{SS})^2\!\! +\!\! {N\over
3}(m^2_0 -\alpha m^2_\pi)^2
\over
            (m^2_\pi -M^2_{SS} +{N\over 3}(\alpha m^2_\pi -m^2_0))^2}\ ,
\end{equation}
or, after the expansion in $m^2_\pi /m^2_{\eta'}$, 
\begin{equation}
I=\CM^2 -Am^2_\pi  
+\left( \CM^2
-2Am^2_\pi
\right) \log{m^2_\pi \over \Lambda^2_{p}}\ , \label{I}
\end{equation}
and 
\begin{eqnarray}
&&\CM^2 \equiv {M^2_{SS} \ m^2_0 \over {M^2_{SS} +Nm^2_0 /3}}
={3y\over 1+Ny}M_{SS}^2  \ , \\
&&A\equiv{\alpha M^4_{SS}+Nm^4_0/3\over\left(M^2_{SS}+N
m^2_0/3\right)^2}
={{\alpha+3Ny^2}\over{(1+Ny)^2}}\ ,
\end{eqnarray}
where $y=(m^2_0/3)/M^2_{SS}$. 

We will examine the  coefficients of the 
chiral logarithm in eq.~(\ref{I}), which is particularly sensitive to 
the chiral limit, in more detail. 
{}From (partially) quenched
lattice data, it is estimated that $m_0^2/3$ presumably has a value
$m_K^2/2\;{<\atop\sim}\;m_0^2/3\;{<\atop\sim}\;m_K^2$ ($m_K=496$~MeV
is the physical kaon mass)
\cite{Sharpem0,delta}. Typical lattice computations have $N=2$ and
$m_K^2\;{<\atop\sim}\;M_{SS}^2\;{<\atop\sim}\;2m_K^2$. These values of
$m_0^2$ and $M_{SS}^2$ correspond to $y$ ranging from $y\approx 1/4$ to
$y\approx 1$.
This leads to ${\cal M}^2/M_{SS}^2=1/2$ for $y=1/4$ and to
${\cal M}^2/M_{SS}^2=1$ for $y=1$.  For $y\rightarrow\infty$ one obtains
${\cal M}^2/M_{SS}^2=3/2$. This shows that for relatively heavy sea
quarks,
there is a clear dependence of the coefficient of the chiral logarithms
on $m_0^2$.
(Experience with quenched lattice data \cite{Sharpem0} indicates that
it is hard to fit the chiral logarithms reliably,
partially because of the ``competition" of $O(p^4)$ coefficients.
This may make it difficult to see the $y$ dependence of
the chiral logarithms in practice.)

The quantity $A$ also has an effect on the coefficients of the chiral
logarithms, in particular for values of the valence-quark mass of order
of the sea-quark mass.
Taking again $N=2$, we find $A=(8\alpha+3)/18$ for $y=1/4$ and
$A=(\alpha+6)/9$ for $y=1$, while $A=3/2$ for $y\rightarrow\infty$.  It
is
clear that $A$ is more sensitive to the value of $\alpha$ for smaller
values
of $y$.   We note that for $m/m_S=1$, our expansion 
parameter $m_\pi^2/m_{\eta'}^2=1/(1+Ny)$
(for $\alpha=0$), so that our results may not be reliable
for smaller values of $y$.  In that case, expression~(\ref{full})
should be used.

\smallskip
We would like to thank Claude Bernard and Steve Sharpe for
discussions.  This work was supported in part by the DOE.

\end{document}